\documentclass[aip,jcp,showpacs,showkeys,preprint,eqsecnum,longbibliography]{revtex4-2}
\usepackage{graphicx}
\usepackage{color}
\usepackage[normalem]{ulem} 

\newcommand{\bq}{\begin{eqnarray}}
\newcommand{\eq}{\end{eqnarray}}
\newcommand{\bqn}{\begin{eqnarray*}}
\newcommand{\eqn}{\end{eqnarray*}}

\newcommand{\rr}{\mathbf{r}}
\newcommand{\RR}{\mathbf{R}}
\newcommand{\xx}{\mathbf{x}}
\newcommand{\yy}{\mathbf{y}}
\newcommand{\zz}{\mathbf{z}}

\newcommand{\qq}{\mathbf{q}}
\newcommand{\QQ}{\mathbf{Q}}

\newcommand{\calp}{{\cal P}}
\newcommand{\calh}{{\cal H}}
\newcommand{\calt}{{\cal T}}
\newcommand{\calv}{{\cal V}}
\newcommand{\cala}{{\cal A}}
\newcommand{\calo}{{\cal O}}
\newcommand{\cald}{{\cal D}}
\newcommand{\calr}{{\cal R}}

\begin{document}
%%%%%%%%%%%%%%%%%%%%%%%%%%%%%%%%%%%%%%%%%%%%%%%%%%%%%%%%%%%%%%%%%%%%%%%%%%%%%%
%%%%%%%%%%%%%%%%%%%%%%%%%%%%%%%%%%%%%%%%%%%%%%%%%%%%%%%%%%%%%%%%%%%%%%%%%%%%%%
%%%%%%%%%%%%%%%%%%%%%%%%%%%%%%%%%%%%%%%%%%%%%%%%%%%%%%%%%%%%%%%%%%%%%%%%%%%%%%
\title{One-component fermion plasma on a sphere at finite temperature. The anisotropy in the 
paths conformations} 

\author{Riccardo Fantoni}
\email{rfantoni@ts.infn.it}
\affiliation{Universit\`a di Trieste, Dipartimento di Fisica, strada
  Costiera 11, 34151 Grignano (Trieste), Italy}
\date{\today}

\begin{abstract}
In our previous work [R. Fantoni, Int. J. Mod. Phys. C, {\bf 29}, 1850064 (2018)] we studied, 
through a computer experiment, a one-component fermion plasma on a sphere at finite, non-
zero, temperature. We extracted thermodynamic properties like the kinetic and internal energy 
per particle and structural properties like the radial distribution function, and produced 
some snapshots of the paths to study their shapes.

Here we revisit such a study giving some more theoretical details explaining the paths shape 
anisotropic conformation due to the inhomogeneity in the polar angle of the variance of the 
random walk diffusion from the kinetic action. 
\end{abstract}


%%\pacs{02.70.Ss,05.10.Ln,05.30.Fk,05.70.-a,61.20.Ja,61.20.Ne}

\maketitle
%%%%%%%%%%%%%%%%%%%%%%%%%%%%%%%%%%%%%%%%%%%%%%%%%%%%%%%%%%%%%%%%%%%%%%%%%%%%%%
\section{Introduction}
%%%%%%%%%%%%%%%%%%%%%%%%%%%%%%%%%%%%%%%%%%%%%%%%%%%%%%%%%%%%%%%%%%%%%%%%%%%%%%
\label{sec:introduction}

In our work of Ref. \cite{Fantoni2018c} we studied, through restricted path integral Monte 
Carlo, a one-component fermion plasma on a sphere of radius $a$ at finite, non-zero, absolute 
temperature $T$. We extracted thermodynamic properties like the kinetic and internal energy 
per particle and structural properties like the radial distribution function, and produced 
some snapshots of the paths to study their shapes.

Our results extended to the quantum regime the previous non-quantum
results obtained for the analytically exactly solvable plasma on
curved surfaces 
\cite{Fantoni03jsp,Fantoni2008,Fantoni2012,Fantoni2012b,Fantoni2016,Fantoni17b}
and for its numerical Monte Carlo experiment \cite{Fantoni12c}. In particular we showed how 
the configuration space (see Fig. 1 of Ref. \cite{Fantoni2018c}) appears much more 
complicated than in the classical case (see Figs. 5 and 6 of Ref. \cite{Fantoni12c}). A first 
notable phenomena is the fact that whereas the particles distribution is certainly isotropic
the paths conformation is not. Some paths tend to wind around the sphere running along the 
parallels in proximity of the poles others to run along the meridians in proximity of the 
equator. This is a direct consequence of the coordinate dependence of the variance of the 
diffusion. At high temperature, the paths tend to be localized, whereas at low temperature, 
they tend to be delocalized distributed over a larger part of the surface with long links 
between the beads.

The plasma is an ensemble of point-wise electrons which interact through the Coulomb 
potential assuming that the electric field lines can permeate the three-dimensional space 
where the sphere is embedded. The system of particles is thermodynamically stable even if the 
pair-potential is purely repulsive because the particles are confined to the compact surface 
of the sphere, and we do not need to add a uniform neutralizing background as in the Wigner
{\sl Jellium} model \cite{Fantoni95b,Fantoni13g,Fantoni21b,Fantoni21d,Fantoni21i}. Therefore 
our spherical plasma made of $N$ spinless indistinguishable electrons of charge $-e$ and mass 
$m$ will carry a total negative charge $-Ne$, a total mass $Nm$, and will have a radius $a$.

In this work we do a thought computer experiment as the one actually carried out in Ref. 
\cite{Fantoni2018c} in order to be able to extract some theoretical conclusions on the paths 
shape and conformation that will try to explain the results found in Ref. \cite{Fantoni2018c}

Our study could be used to predict the properties of a metallic
spherical shell, as for example a spherical shell of graphene. Today we assisted the
rapid development of the laboratory realization of graphene hollow spheres 
\cite{Rashid,Tiwari} with
many technological interests like the employment as electrodes for supercapacitors
and batteries, as superparamagnetic materials, as electrocatalysts for oxygen reduction, as 
drug deliverers, as a conductive catalyst for photovoltaic applications 
\cite{Guo2010,Cao2013,Wu2013,Shao2013,Zhao2016,Cho2016,Hao2016,Huang2017,Chen2017}. Of
course, with simulation we can access the more various and extreme conditions
otherwise not accessible in a laboratory.

A possible further study would be the simulation of the neutral sphere
where we model the plasma of electrons as embedded in a spherical
shell that is uniformly positively charged in such a way that the
system is globally neutrally charged. This can easily be done by
changing the Coulomb pair-potential into $e^2/r\to e^2(1/r-1)$. In the
$a\to\infty$ limit, this would reduce to the Wigner Jellium model
which has been received much attention lately, from the point of view
of a path integral Monte Carlo simulation
\cite{Brown2013,Brown2014,Dornheim2016,Dornheim2016b,Groth2016,Groth2017,Malone2016,Filinov2015,Fantoni2018c,Fantoni2018b}
. Alternatively
we could study the two-component plasma on the sphere as has recently
been done in the tridimensional Euclidean space \cite{Fantoni2018b}. 
Another 
possible extension of our work is the realization of the simulation of
the full anyonic plasma on the sphere taking care appropriately of the
fractional statistics and the phase factors to append to each
disconnected region of the path integral expression for the partition
function \cite{Fantoni2018c}. This could become important in a study
of the quantum Hall effect by placing a magnetic Dirac monopole at the
center of the sphere \cite{Melik1997,Melik2001}. Also the adaptation
of our study to a fully relativistic Hamiltonian could be of some
interest for the treatment of the Dirac points graphinos.

The paper is organized as follows: in section \ref{sec:problem} we
describe the thought system and the method used for its study, 
in section \ref{sec:results} we present our theoretical study and predictions, 
and section \ref{sec:conclusions} is for the concluding discussion. 

%%%%%%%%%%%%%%%%%%%%%%%%%%%%%%%%%%%%%%%%%%%%%%%%%%%%%%%%%%%%%%%%%%%%%%%%%%%%%%
\section{The problem}
%%%%%%%%%%%%%%%%%%%%%%%%%%%%%%%%%%%%%%%%%%%%%%%%%%%%%%%%%%%%%%%%%%%%%%%%%%%%%%
\label{sec:problem}

A point $\qq$ on the sphere of radius $a$, the surface of constant
positive curvature, is given by 
\bq
\rr/a=\sin\theta\cos\varphi\hat{\xx}+\sin\theta\sin\varphi\hat{\yy}+
\cos\theta\hat{\zz},
\eq
with $\theta$ the polar angle and $\varphi$ the azimuthal angle.
The $N$ particles positions are at
$\RR=(\rr_1,\rr_2,\ldots,\rr_N)$. The surface density of the plasma
will then be $\sigma=N/4\pi a^2$.
On the sphere we have the following metric
\bq \label{metric}
ds^2=g_{\mu\nu}dq^\mu dq^\nu=a^2\left[d\theta^2+\sin^2\theta d\varphi^2\right],
\eq 
where Einstein summation convention on repeated indices is assumed, we
will use greek indices for either the surface components or the surface
components of each particle coordinate and roman indices for either
the particle index or the time-slice index, $q^1=\theta\in
[0,\pi)$, $q^2=\varphi\in [-\pi,\pi)$, and the positive definite
and symmetric metric tensor is given by
\bq
g_{\mu\nu}=\left(\begin{array}{cc}
a^2 & 0 \\
0   & a^2\sin^2\theta
\end{array}\right).
\eq
We have periodic boundary conditions in $\theta+\pi=\theta$ and in
$\varphi+2\pi=\varphi$. We will also define 
$\QQ=(\qq_1,\qq_2,\ldots,\qq_N)$. The geodesic distance between two 
infinitesimally close points $\QQ$ and $\QQ'$ is
$ds^2(\QQ,\QQ')=\sum_{i=1}^Nds^2(\qq_i,\qq_i')$ where the geodesic
distance between the points $\qq$ and $\qq'$ on the sphere is
\bq \nonumber
s(\qq,\qq')&=&a\arccos\left[\cos(q^1)\cos({q^1}')+\right.\\ \label{gd}
&&\left.\sin(q^1)\sin({q^1}')\cos(q^2-{q^2}')\right].
\eq

The Hamiltonian of the $N$ non-relativistic indistinguishable particles
of the one-component spinless fermion plasma is given by 
\bq \label{Ham}
\calh=\calt+\calv=-\lambda\sum_{i=1}^N\Delta_i+\sum_{i<j}v_{ij},
\eq
with $\lambda=\hbar^2/2m$, where $m$ is the electron mass, and
$\Delta_i=g_i^{-1/2}\partial(g_i^{1/2} g_i^{\mu\nu}\partial/\partial
q_i^\nu)/\partial q_i^\mu$ the Laplace-Beltrami operator for
the $i$th particle on the sphere of radius $a$ in local coordinates,
where $g_{\mu\alpha}g^{\alpha\nu}=\delta_\mu^\nu$ and $g_i=\det
||g_{\mu\nu}(\qq_i)||$. We have assumed that 
$\calh$ in curved space has the same form as in flat space. For the
pair-potential, $v$, we will choose 
\bq
v_{ij}=e^2/r_{ij},
\eq
where $e$ is the electron charge and $r_{ij}$ is the Euclidean
distance between two particles at $\qq_i$ and $\qq_j$, which is given
by 
\bq
r_{ij}=a\sqrt{2-2\hat{\rr}_i\cdot\hat{\rr}_j}=
2a\sin[\arccos(\hat{\rr}_i\cdot\hat{\rr}_j)/2],
\eq
where $\hat{\rr}_i=\rr_i/a$ is the versor that from the center of the
sphere points towards the center of the $i$th particle. So the electrons 
{\sl move on} a spherical shell with the electric field lines permeating 
the surrounding three dimensional space, they do not {\sl live in} the shell.

Given the antisymmetrization operator $\cala=\sum_P/N!$, where the sum runs over all 
particles permutations $P$, and the inverse temperature $\beta=1/k_BT$, with $k_B$ 
Boltzmann's constant, the one-component fermion plasma density matrix, $\rho_F=\cala
e^{-\beta\calh}$, in the coordinate representation, on a generic
Riemannian manifold of metric $g$ \cite{Fantoni12e,Schulman}, is
\begin{widetext}
\bq \nonumber
\rho_F({\QQ^\prime},\QQ;\beta)=\int&&
\rho_F({\QQ^\prime},\QQ((M-1)\tau);\tau)
\cdots
\rho_F(\QQ(\tau),\QQ;\tau)\times\\ \label{rhof}
&&\prod_{j=1}^{M-1}\sqrt{\tilde{g}_{(j)}}\prod_{i=1}^N\,
dq_i^1(j\tau)\wedge dq_i^2(j\tau)~,
\eq
where as usual we discretize the {\sl imaginary thermal time} in bits
$\tau=\hbar\beta/M$. We will often use the following shorthand
notation for the {\sl path integral} measure:
$\prod_{j=1}^{M-1}\sqrt{\tilde{g}_{(j)}}\prod_{i=1}^N\,$ 
$dq_i^1(j\tau)\wedge dq_i^2(j\tau)\to\cald\QQ$ as $M\to\infty$. The
path of the $i$th particle is given by
$\{\qq_i(t)|t\in[0,\hbar\beta]\}$ with $t$ the 
imaginary thermal time. Each $\qq_i(j\tau)$ with $i=1,\ldots,N$ and
$j=1,\ldots,M$ represents the various {\sl beads} forming the
discretized path. The $N$ particle path is given by
$\{\QQ(t)|t\in[0,\hbar\beta]\}$. Moreover,  
\bq
\tilde{g}_{(j)}&=&\det||\tilde{g}_{\mu\nu}(\QQ(j\tau))||,~~~j=1,2,\ldots,M-1,\\
\tilde{g}_{\mu\nu}(\QQ)&=&g_{\alpha_1\beta_1}(\qq_1)\otimes\ldots
\otimes g_{\alpha_N\beta_N}(\qq_N),
\eq
In the small $\tau$ limit we have 
\bq \nonumber
\rho_F(\QQ(2\tau),\QQ(\tau);\tau)\propto \cala\left[
\tilde{g}_{(2)}^{-1/4}\sqrt{D(\QQ(2\tau),\QQ(\tau);\tau)}
\tilde{g}_{(1)}^{-1/4}\times\right.\\
\left.e^{\lambda\tau R(\QQ(\tau))/6\hbar}
e^{-\frac{1}{\hbar}S(\QQ(2\tau),\QQ(\tau);\tau)}\right],
\eq
\end{widetext}
where $\cala$ can act on the first, or on the second, or on both {\sl
time slices}, $R(\QQ)$ the scalar curvature of the curved manifold,
$S$ the action and $D$ the van Vleck's determinant  
\bq
D_{\mu\nu}&=&\frac{\partial^2S(\QQ(2\tau),\QQ(\tau);\tau)}
{\partial {Q}^\mu(2\tau)\partial {Q}^\nu(\tau)},\\
\det||D_{\mu\nu}||&=&D(\QQ(2\tau),\QQ(\tau);\tau),
\eq
where here the greek index denotes the two components of each particle
coordinate.  

For the {\sl action} and the {\sl kinetic-action} we have 
\bq
S(\QQ',\QQ)&=&K(\QQ',\QQ)+U(\QQ',\QQ),\\
K(\QQ',\QQ)&=&N\hbar\ln(4\pi\lambda\tau/\hbar)+
\frac{\hbar^2s^2(\QQ',\QQ)}{4\lambda\tau},
\eq
where in the {\sl primitive approximation} \cite{Ceperley1995} we find
the following expression for the {\sl inter-action}, 
\bq \label{primitive}
U(\QQ',\QQ)&=&\frac{\tau}{2}[V(\QQ')+V(\QQ)],\\
V(\QQ)&=&\sum_{i<j}v_{ij}.
\eq
In particular the kinetic-action is responsible for a diffusion of the
random walk with a variance of $2\lambda\tau g^{\mu\nu}/\hbar$.

On the sphere we have $R=N\calr$ with $\calr=2/a^2$, the scalar
curvature of the sphere of radius $a$, and in the $M\to\infty$ limit 
$s^2(\QQ',\QQ)\to ds^2(\QQ',\QQ)$ and
$\tilde{g}_{(2)}^{-1/4}$ $\sqrt{D(\QQ(2\tau),\QQ(\tau);\tau)}$ 
$\tilde{g}_{(1)}^{-1/4}\to\left(\hbar^2/2\lambda\tau\right)^{N}$.
\footnote{For a space of constant curvature there is clearly no effect, as the term due to 
the curvature just leads to a constant multiplicative factor that has no influence on the 
measure of the various observables. One might have hoped that certain constrained 
coordinates, perhaps a relative coordinate in a molecule, would effectively live in a space 
of variable curvature. Perhaps gravitation will give us the system on which the effect of 
curvature can be seen, but at present the effect is purely in the realm of theory.} We
recover the Feynman-Kac path integral formula on the sphere in the
$\tau\to 0$ limit. We will then have to deal with $2NM$
multidimensional integrals for which Monte Carlo \cite{Kalos-Whitlock} is a suitable
computational method. For example to measure an observable $\calo$ we
need to calculate the following quantity
\bq
\langle\calo\rangle=\frac{\int
  O(\QQ,\QQ')\rho_F(\QQ',\QQ;\beta)\,d\QQ d\QQ'}
{\int \rho_F(\QQ,\QQ;\beta)\,d\QQ},
\eq
where $\sqrt{\tilde{g}}\prod_{i=1}^N\,dq_i^1\wedge dq_i^2\equiv
d\QQ$. Notice that most of the properties 
that we will measure are diagonal in coordinate representation,
requiring then just the diagonal density matrix,
$\rho_F(\QQ,\QQ;\beta)$.

For example for the density $\rho(\qq)=\langle\calo\rangle$ with 
\bq \label{density}
O(\QQ;\qq)=\sum_{i=1}^N\delta^{(2)}(\qq-\qq_i),
\eq
where $\delta^{(2)}(\qq)=\delta(q^1)\delta(q^2)$ and $\delta$ is the Dirac delta function.
Clearly $\int\sigma(\qq)\,\sqrt{g(\qq)}d\qq=N$ and a uniform distribution of electrons 
is signaled by a constant density throughout the surface of the sphere. 

Fermions' properties cannot be calculated exactly with path integral
Monte Carlo because of the fermions sign problem
\cite{Ceperley1991,Ceperley1996}. We then have to resort to an
approximated calculation. The one we chose in Ref. \cite{Fantoni2018c} 
was the restricted path integral approximation \cite{Ceperley1991,Ceperley1996} 
with a ``free fermions restriction''. The trial density matrix used in the
restriction is chosen as the one reducing to the ideal density matrix
in the limit of $t\ll 1$ and is given by
\bq \label{ifdm}
\rho_0(\QQ',\QQ;t)\propto\cala \left|\left|e^{
-\frac{\hbar s^2(\qq_i',\qq_j)}
{4\lambda t}}\right|\right|.
\eq
The {\sl restricted path integral identity} that can be used
\cite{Ceperley1991,Ceperley1996} is as follows 
\bq \label{rpii}
\rho_F(\QQ',\QQ;\beta)&\propto&\int \sqrt{\tilde{g}''}d\QQ''\,\rho_F(\QQ'',\QQ;0)\times
\\ \nonumber
&&\oint_{\QQ''\to\QQ'\in\gamma_0(\QQ)}\cald\QQ'''\,e^{-S[\QQ''']/\hbar},
\eq
where $S$ is the Feynman-Kac action
\bq
S[\QQ]=\int_0^{\hbar\beta} dt\left[\frac{\hbar^2}{4\lambda}
\dot{\QQ}_\mu\dot{\QQ}^\mu+V(\QQ)\right],
\eq
here the dot indicates a total derivative with respect to the
imaginary thermal time, and the subscript in the path integral of
Eq. (\ref{rpii}) means that we restrict the path integration to 
paths starting at $\QQ''$, ending at $\QQ'$ and avoiding the nodes of
$\rho_0$, that is to the {\sl reach} of $\QQ$, $\gamma_0$. The nodes
are on the reach boundary $\partial\gamma_0$. The weight of
the walk is $\rho_F(\QQ'',\QQ;0)=\cala\delta(\QQ''-\QQ)$
$=(N!)^{-1}\sum_\calp(-)^\calp$ $\delta^{(2N)}(\QQ''-\calp\QQ)$, where the
sum is over all the permutations $\calp$ of the $N$ fermions,
$(-)^\calp$ is the permutation sign, positive for an even permutation
and negative for an odd permutation, and the $\delta$ is a Dirac delta function. 
It is clear that the  
contribution of all the paths for a single element of the density
matrix will be of the same sign, thus solving the sign problem;
positive if $\rho_F(\QQ'',\QQ;0)>0$, negative otherwise. On the
diagonal the density matrix is positive and on the path restriction
$\rho_F(\QQ',\QQ;\beta)>0$ then only even permutations are allowed
since $\rho_F(\QQ,\calp\QQ;\beta)=(-)^\calp\rho_F(\QQ,\QQ;\beta)$. It
is then possible to use a bosons calculation to get the fermions
case. Clearly the restricted path integral identity with the free
fermions restriction becomes exact if we simulate free fermions, but
otherwise is just an approximation. The approximation is expected to
become better at low density and high temperature, i.e. when
correlation effects are weak. The implementation of the restricted,
fixed nodes, path integral identity within the worm algorithm has been
also the subject of our previous study on the three-dimensional Euclidean 
Jellium \cite{Fantoni21b}.  

In Ref. \cite{Fantoni2018c} we worked in the grand canonical ensemble 
with fixed chemical
potential $\mu$, surface area $A=4\pi a^2$, and absolute temperature
$T$. At a higher value of the chemical potential we will have a higher
number of particles on the surface and a higher density. On the other
hand, increasing the radius of the sphere at constant chemical
potential will produce a plasma with lower surface density.  
The {\sl Coulomb coupling constant} is $\Gamma=\beta e^2/a_0r_s$ with
$a_0=\hbar^2/me^2$ the Bohr radius and
$r_s=(4\pi\sigma)^{-1/2}/a_0$. At weak coupling, $\Gamma\ll 1$, the 
plasma becomes weakly correlated and approaches the ideal gas
limit. This will occur at high temperature and/or low density. The
{\sl electron degeneracy parameter} is $\Theta=T/T_D$ where the
degeneracy temperature $T_D=\sigma\hbar^2/mk_B$. For temperatures
higher than $T_D$, $\Theta\gg 1$, quantum effects are less relevant.

%%%%%%%%%%%%%%%%%%%%%%%%%%%%%%%%%%%%%%%%%%%%%%%%%%%%%%%%%%%%%%%%%%%%%%%%%%%%%%
\section{Theoretical study}
%%%%%%%%%%%%%%%%%%%%%%%%%%%%%%%%%%%%%%%%%%%%%%%%%%%%%%%%%%%%%%%%%%%%%%%%%%%%%%
\label{sec:results}

In order to understand the anisotropic conformation of the paths snapshots and their 
dependence on the azimuthal angle $\varphi$ and polar angle $\theta$ we observe that in the 
primitive approximation we have in the path integral a weight factor 
$\propto \exp(-\hbar ds^2/4\lambda\tau)$ stemming from the kinetic part of the action, where 
$ds^2$ is given by Eq. (\ref{metric}). In particular we see that if we are near the 
poles, $\theta\approx 0$ or $\pi$, then $ds^2\approx a^2d\theta^2$ and we see that it costs 
nothing to change the azimuthal angle. This explains the paths winding along the parallels in 
proximity of the poles. On the other hand near the equator, at $\theta\approx\pi/2$, we find 
$ds^2\approx a^2(d\theta^2+d\varphi^2)$ so that the paths will tend to wonder around the 
equator in no particular direction. 

\begin{figure}[htbp]
\begin{center}
\includegraphics[width=10cm]{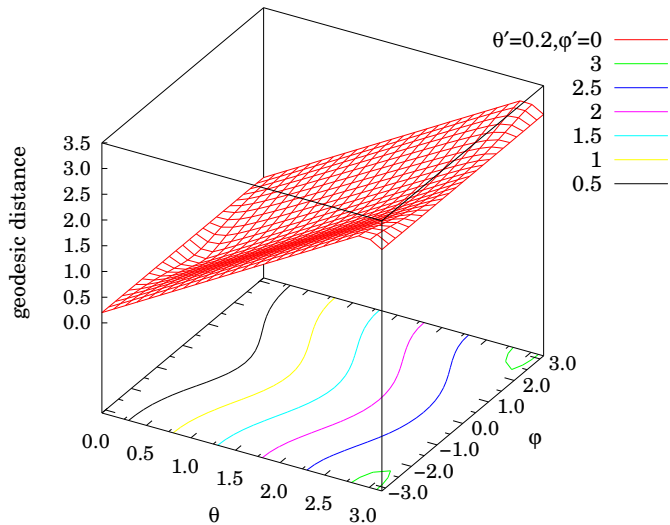}
\end{center}  
\caption{(color online) Three dimensional plot of the geodesic distance of Eq. (\ref{gd}) for 
$\theta '=0.2$ and $\varphi '=0$. From the surface graph we see how in proximity of the poles 
the geodesic distance between points moving along parallels is small while it increases 
rapidly if one moves along the meridians.}
\label{fig:gd1}
\end{figure}

\begin{figure}[htbp]
\begin{center}
\includegraphics[width=10cm]{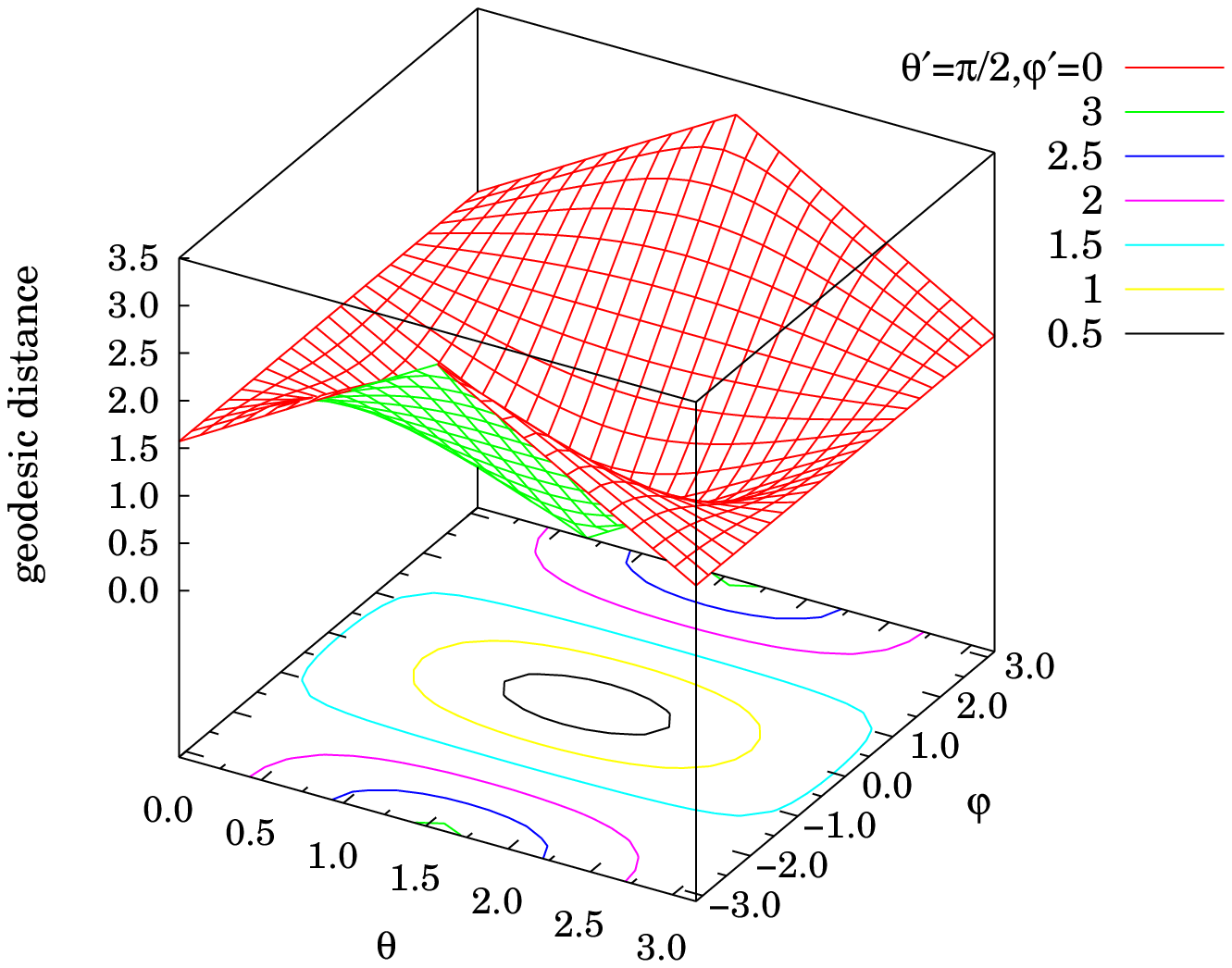}
\end{center}  
\caption{(color online) Three dimensional plot of the geodesic distance of Eq. (\ref{gd}) for 
$\theta '=\pi/2$ and $\varphi '=0$. From the surface graph we see how in proximity of the 
equator the geodesic distance between points moving in circles in the $(\theta,\varphi)$ 
plane is small while it increases rapidly if one moves along the parallels.}
\label{fig:gd2}
\end{figure}

The same can be seen studying the behavior of the finite geodesic distance of Eq. (\ref{gd}). 
In Fig. (\ref{fig:gd1}) we show a three dimensional plot for $\theta '=0.2$ and 
$\varphi '=0$. Again we see that around the pole at $\theta\approx 0$ it costs nothing to 
change $\varphi$, that is to go along a parallel, while a path traveling along a meridian 
will be unfavored since we need to increase $\theta$. In Fig. (\ref{fig:gd2}) we show a three 
dimensional plot for $\theta '=\pi/2$ and $\varphi '=0$. And now we see that around the 
equator at $\theta\approx\pi/2$ it is favored a path wondering around the initial position 
with no preferred direction along the parallels or the meridians.


Clearly if we rotate the sphere the paths shape will simply rotate following the rotation of 
her poles. This anisotropy of the path conformations is rather counter intuitive since the 
sphere is notoriously isotropic but it reflects the inhomogeneity of the metric respect to 
the polar angle.

It is important to distinguish the effect that we just described due to the weight factor 
$\propto \exp(-\hbar ds^2/4\lambda\tau)$ stemming from the kinetic part of the primitive 
action from the measure factor $\prod_{j=1}^{M}\sqrt{\tilde{g}(j)}$ also entering the path 
integral. This last factor being also independent of the azimuthal angles will produce the 
same local density $\rho(\qq)$ under a rotation of the sphere around her axis through the 
poles. So, by isotropy, we conclude that the density must be a constant under any rotation, 
which means that the plasma must be uniform \cite{Fantoni12e}.

The temperature dependence can also be easily explained. at high temperature $\beta$ is 
small, $\Theta\gg 1$, and the path extending from $t=0$ to $t=\hbar\beta$ will be localized, 
of a small size, and quantum effects will be less relevant, whereas at low temperature 
$\beta$ is large, $\Theta\ll 1$, and the path will be delocalized, increased in size, it 
diffuses more on the surface, and quantum effects are more relevant. Usually we will be 
interested in measuring observables which are diagonal so that when dealing with the diagonal 
density matrix $\rho_F(\QQ,\QQ;\beta)$ we will observe {\sl ring} paths such that 
$\QQ(0)=\QQ(\hbar\beta)$. Moreover at high 
temperature the diagonal density matrix will involve almost straight localized ring paths 
closing themselves on the identity permutation. Whereas at low temperature the delocalized 
paths will eventually wind through the $\hbar\beta$ periodicity by means of several different 
permutations $\calp$, so that $\QQ(0)=\calp\QQ(\hbar\beta)$ and so on. 
Any permutation can be broken into a product of cyclic permutations. Each cycle corresponds 
to several paths ``cross-linking'' and forming a larger ring path. Quantum mechanically the 
plasma does this to lower its kinetic energy. According to Feynman's 1953 theory 
\cite{Ceperley1995}, the superconductor transition is represented by the formation of 
macroscopic paths, i.e., those stretching across the whole sphere and involving on the order 
of N electrons. Or in other words, those ring paths percolating through the periodic boundary 
conditions $\theta=\theta+\pi$ and $\varphi=\varphi+2\pi$ by means of permutations. 

\begin{figure}[htbp]
\begin{center}
\includegraphics[width=12cm]{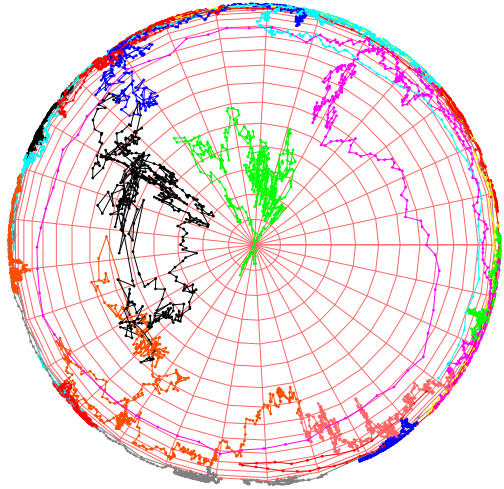}\\[-3cm]
\includegraphics[width=12cm]{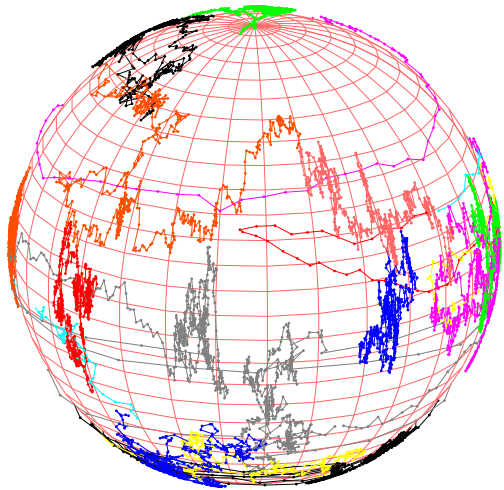}
\end{center}  
\caption{(color online) Snapshot of the macroscopic path during a simulation. 
The different paths have different colors. The simulation parameter specifying the 
thermodynamic condition are as follows: $a=5, \lambda=1, \beta=5, \mu=8$.}
\label{fig:snapshot}
\end{figure}

In Fig. \ref{fig:snapshot} we show a snapshot of the macroscopic path during a simulation of 
Ref. \cite{Fantoni2018c}.
% 
%%%%%%%%%%%%%%%%%%%%%%%%%%%%%%%%%%%%%%%%%%%%%%%%%%%%%%%%%%%%%%%%%%%%%%%%%%%%%%
\section{Conclusions}
%%%%%%%%%%%%%%%%%%%%%%%%%%%%%%%%%%%%%%%%%%%%%%%%%%%%%%%%%%%%%%%%%%%%%%%%%%%%%%
\label{sec:conclusions}

In this work we revised our restricted path integral Monte Carlo
simulation \cite{Fantoni2018c} of a one-component spinless fermion plasma at finite,
non-zero, temperature on the surface of a sphere. The Coulomb
interaction is $e^2/r$ with $r$ the Euclidean distance between the two
electrons of elementary charge $e$ (we could as well have chosen
instead of $r$ the geodesic distance, $s$, within the sphere). This gives 
us an approximated numerical solution of the many-body problem. The
exact solution cannot be accessed due to the fermion sign
catastrophe. Impenetrable indistinguishable particles on the surface
of a sphere admit, in general, anyonic statistics \cite{Fantoni21a}. 
Here we just project the larger bride group onto the permutation group 
and choose the fermion sector for our study.

The path integral Monte Carlo method chosen in Ref. \cite{Fantoni2018c} 
used the primitive approximation for the action which could be improved 
for example by the use of the pair-product action \cite{Ceperley1995}. The
restriction was carried on choosing as the trial density matrix the one
of ideal free fermion. This choice would of course return an exact solution for
the simulation of ideal fermions but it furnishes just an approximation
for the interacting coulombic plasma.

In this work we showed how the conformation anisotropy of the paths observed in the 
simulations of Ref. \cite{Fantoni2018c} can be explained through the inhomogeneous nature of 
the metric in the polar angle. Or equivalently from the inhomogeneous nature of the geodesic 
distance on the surface of the sphere. And this is ultimately due to the fact that the metric 
enters with the negative sign in the exponent of the primitive approximation of the density 
matrix. We should not confuse the anisotropy in the paths conformation with the fact that the 
plasma will always be homogeneous (with a constant local density $\rho$) on the sphere. 
In the degenerate regime (low $T$) the observed strong anisotropy in the path conformation 
near the poles or the equator of the sphere should also be due to a peculiar behavior in the
properties of the $N$-particle off-diagonal density matrix. This, as it is well known, 
is directly related with a number of physical properties, like the quasi-particle excitation
spectrum and the momentum distribution. Therefore, the system properties
can deviate significantly from just a pure homogeneous 2D system, and the
inhomogeneous nature of the space metric is of a particular importance.

We also suggested the possibility to observe a superconducting plasma at low 
temperature when we observe ring paths percolating through the periodic boundary conditions 
$\theta=\theta+\pi$ and $\varphi=\varphi+2\pi$ by means of permutations, even if some care 
has to be addressed to take into account the peculiar asymptotic behavior of the one-particle 
density matrix. 

%%%%%%%%%%%%%%%%%%%%%%%%%%%%%%%%%%%%%%%%%%%%%%%%%%%%%%%%%%%%%%%%%%%%%%%%%%%%%% 
%\begin{acknowledgments}
%...
%\end{acknowledgments} 
%%%%%%%%%%%%%%%%%%%%%%%%%%%%%%%%%%%%%%%%%%%%%%%%%%%%%%%%%%%%%%%%%%%%%%%%%%%%%%
%\bibliographystyle{}
\bibliography{jft}

%%%%%%%%%%%%%%%%%%%%%%%%%%%%%%%%%%%%%%%%%%%%%%%%%%%%%%%%%%%%%%%%%%%%%%%%%%%%%%
%%%%%%%%%%%%%%%%%%%%%%%%%%%%%%%%%%%%%%%%%%%%%%%%%%%%%%%%%%%%%%%%%%%%%%%%%%%%%%
%%%%%%%%%%%%%%%%%%%%%%%%%%%%%%%%%%%%%%%%%%%%%%%%%%%%%%%%%%%%%%%%%%%%%%%%%%%%%%
\end{document}